\documentclass[conference]{IEEEtran}
\IEEEoverridecommandlockouts

\usepackage{cite}
\usepackage{amsmath,amssymb,amsfonts}
\usepackage{algorithmic}
\usepackage{graphicx}
\usepackage{textcomp}
\usepackage{xcolor}
\usepackage{url}
\def\BibTeX{{\rm B\kern-.05em{\sc i\kern-.025em b}\kern-.08em
    T\kern-.1667em\lower.7ex\hbox{E}\kern-.125emX}}
\begin{document}

\title{Improving Pronunciation and Accent Conversion through Knowledge Distillation And Synthetic Ground-Truth from Native TTS\\

}

\author{
   \vspace{-0.5cm}
    \IEEEauthorblockN{
        Tuan Nam Nguyen\IEEEauthorrefmark{1}, 
        Seymanur Akti\IEEEauthorrefmark{2}, 
        Ngoc Quan Pham\IEEEauthorrefmark{3}, 
        Alexander Waibel\IEEEauthorrefmark{4}
    }
    \\
    \IEEEauthorblockA{
        \IEEEauthorrefmark{1}\IEEEauthorrefmark{2}\IEEEauthorrefmark{3}\IEEEauthorrefmark{4}
        Karlsruhe Institute of Technology, Karlsruhe, Germany
    }
    \IEEEauthorblockA{
        \IEEEauthorrefmark{4} Carnegie Mellon University, Pittsburgh PA, USA
    }
    
    \IEEEauthorblockA{
        \IEEEauthorrefmark{1}tuan.nguyen@kit.edu, 
        \IEEEauthorrefmark{2}seymanur.akti@kit.edu, 
        \IEEEauthorrefmark{3}ngoc.pham@kit.edu, 
        \IEEEauthorrefmark{4}alexander.waibel@kit.edu
    }    
    \vspace{-0.9cm}
}

\maketitle

\begin{abstract}
Previous approaches on accent conversion (AC) mainly aimed at making non-native speech sound more native while maintaining the original content and speaker identity. However, non-native speakers sometimes have pronunciation issues, which can make it difficult for listeners to understand them. Hence, we developed a new AC approach that not only focuses on accent conversion but also improves  pronunciation  of non-native accented speaker. By providing the non-native audio and the corresponding transcript, we generate the ideal ground-truth audio with native-like pronunciation with original duration and prosody. This ground-truth data aids the model in learning a direct mapping between accented and native speech. We utilize the end-to-end VITS framework to achieve high-quality waveform reconstruction for the AC task. As a result, our system not only produces audio that closely resembles native accents and while retaining the original speaker's identity but also improve pronunciation, as demonstrated by evaluation results.
\end{abstract}

\vspace{-0.3cm}
\section{Introduction}
Second-language (L2) English learners typically present accents and mispronunciations, which can greatly impact their communication proficiency.  AC basically aims to change the accent of speech while retaining the information of content and speaker identity. However, this task remains particularly challenging due to the extreme lack of parallel corpora and the large variations in different speakers. This research focuses on investigating techniques to enhance AC models' capacity to improve pronunciation made by L2 speakers and devising appropriate evaluation measures for this purpose. 

Conventional AC \cite{8462258,zhao19f_interspeech} methods require reference utterances in the target accent during synthesis, limiting their applicability due to the challenge of obtaining linguistically identical references in different accents. This research, like recent studies \cite{jia2023zeroshot,jin2022voicepreserving,9477581,huang2023evaluatingmethodsgroundtruthfreeforeign,waibel2012simultaneous}, focuses on reference-free AC, which converts accents without needing reference utterances during inference. Reference-free AC methods generally fall into two categories: autoregressive and non-autoregressive architectures.
    
Autoregressive AC models , typically based on sequence-to-sequence (seq2seq) architectures, rely mostly on parallel data for training. However, such data—utterances from the same speakers in different accents—is scarce, making it difficult to obtain. To address this, data augmentation techniques like text-to-speech (TTS) \cite{10096431} or voice conversion \cite{AC_NamNguyen} are used to generate ground-truth audio for non-native speakers. These ground-truth audios, originating from different speakers, often differ in duration and prosody compared to the original non-native audio. While the encoder-decoder model with attention mechanisms can help aligning inputs and outputs, the differences in length make these models less suitable for applications requiring precise time synchronization, such as video conference dubbing \cite{waibe112005chil,waibel1998meeting, huang2003automatic}. Additionally, there are some autoregressive models\cite{10107770,9053797}  that trained with non-parallel data, but still share similar drawbacks, including slow inference speeds and unstable attention mechanism. 

In contrary, non-autoregressive AC leverages non-parallel data, which is abundant and diverse. To effectively utilize this data, these models are built on a disentangle-resynthesis framework. Here, speech is disentangled into separate  features such as speaker identity, content, prosody, and accent, which are then recombined to generate the original waveform. Content features are often represented by bottleneck features (BNFs), extracted from self-supervised pre-trained models \cite{10095191} (e.g., WavLM \cite{9814838}, Wav2vec \cite{NEURIPS2020_92d1e1eb}, HuBERT \cite{10.1109/TASLP.2021.3122291}) , ASR bottlenecks \cite{jia2023zeroshot}, or ASR output in logits form \cite{10094737}. However, BNFs also capture accent-related feature and non-native pronunciation issues. While methods like adversarial training  \cite{10094737}or Pseudo-Siamese networks  \cite{jia2023zeroshot} attempt to disentangle accent information, they still struggle to improve pronunciations due to the absence of ground-truth.


Building on the strengths and limitations of existing AC models, we propose a novel framework for training a non-autoregressive AC model using generated parallel data. We hypothesize that a TTS system trained solely on native speech will produce accent-independent linguistic representations. Additionally, this native TTS system is able to generate ideal ground-truth data for non-native speakers, ensuring native pronunciation, same speaker identity, duration, prosody, and precise alignment with the original non-native audio. The accent-independent linguistic representations learned by the native TTS model not only facilitate ground-truth generation but also distill knowledge into our AC model, aiding in the learning of accent-independent features. 

Previous studies \cite{10107770,10447205} have used native TTS models to guide AC models, often by sharing the same decoder and distilling knowledge from the TTS text encoder to the AC encoder. A key challenge in these approaches is aligning the outputs of the TTS and AC encoders due to their differing lengths. In \cite{10107770}, both models use the autoregressive Tacotron \cite{8461368} architecture, necessitating an external unstable attention mechanism for alignment. Another approach \cite{10447205} employs the non-autoregressive FastSpeech 2 \cite{ren2021fastspeech} TTS model. In this approach, alignment is achieved through an external length regulator based on Montreal Force Alignment, which uses a GMM-HMM ASR model to upsample the TTS encoder's output, enabling both encoders to produce outputs of similar length. This latter method closely aligns with our proposed approach. These approaches use a separate vocoder to convert mel-spectrograms into waveforms.

The non-autoregressive VITS framework has proven successful in both TTS \cite{kim2021conditional} and Voice Conversion \cite{10095191}. Inspired by this success, we adopt a modified VITS framework for our AC system due to several advantages. VITS uses internal monotonic alignment, simplifying audio-text alignment and improving training efficiency. Additionally, as an end-to-end architecture, VITS eliminates the need for a separate vocoder to convert mel-spectrograms into waveforms, setting it apart from previous AC methods.

Overall, to further enhance the AC model's training process, we introduced several key improvements. In the first stage, we simultaneously pre-train the AC model while training the VITS-based native TTS, allowing the AC model to learn the distribution of native audio data comprehensively.  After training the native VITS-based TTS, we leverage it to generate ideal ground-truth data for non-native audio, as described in the next section. In the second stage, we fine-tune the pre-trained AC model  using the non-native input alongside the generated ground-truth output and also distillate knowledge from the native TTS. In summary, we train the VITS-based TTS model to assist in initializing the weights, generating ideal ground-truth data, and facilitating knowledge distillation for the AC model.{"originatingScript":"m2","payload":{"guid":"9a5f9d28-6c94-4bff-a24e-b5e10ecf90e15b8e9d","muid":"1a8ee153-6a60-4b16-82ea-42e185928d0a78c2dd","sid":"ea8eb107-bf17-4e70-b563-aa64cdb85f85496a83"}}

\vspace{-0.2cm}
\section{Methodology}

\begin{figure*}[t!]
   
  \includegraphics[width=\textwidth]{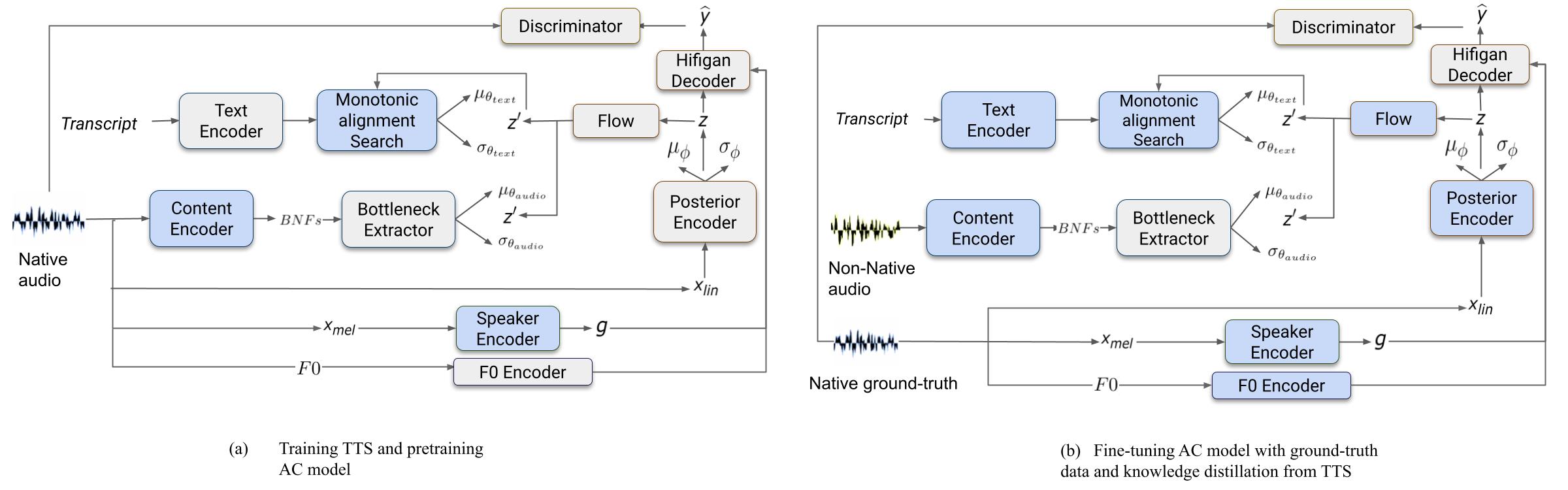}

    \caption{ Pre-training and fine-tuning procedure of our proposed model. The parameters of blue components are frozen.}
  \label{fig:freevc-arch}
\vspace{-5mm}
\end{figure*}
\vspace{-0.1cm}
\subsection{Training native VITS and pretrained AC model}
In the first  stage, we pre-train the AC model and train the native TTS model using the VITS framework. VITS  \cite{kim2021conditional} is a conditional variational autoencoder architecture augmented with normalizing flow \cite{pmlr-v37-rezende15}. It consists of three main components: a posterior encoder, a prior encoder, and a waveform generator. These modules encode the distributions \(q_\phi(z|x)\), \(p_\theta(z|c)\), and \(p_\psi(y|z)\), respectively. Here, \(q_\phi(z|x)\) and \(p_\psi(y|z)\) represent the posterior and data distributions, parameterized by the posterior encoder \(\phi\) and HiFi-GAN \cite{NEURIPS2020_c5d73680} waveform generator \(\psi\), where \(x\) is the speech input,\(z\) is the latent variable and \(y\) is the waveform output. The prior distribution of \(z\) is defined as \(p_\theta(z|c)\),  parameterized by the prior encoder \(\theta\) and refined through a normalizing flow \(f\), where the latent variables are conditioned on the input \(c\), which can be either audio or text.

Our AC and TTS models share the HiFi-GAN decoder, posterior encoder, speaker encoder, F0 encoder and normalizing flow, as shown in Fig.~\ref{fig:freevc-arch}a. The posterior encoder takes linear spectrograms \(x_{lin}\) as input to sample a latent variable \(z\). This latent \(z\), along with speaker embedding \(g\) from speaker encoder and F0 sequence embeddings \(F0\) from F0 encoder, are fed into the HiFi-GAN vocoder to generate the waveform.

The AC and TTS models each have separate prior encoders. The AC model's prior encoder \(\theta_{audio}\) includes a content encoder which extracts speech features and a bottleneck extractor to generate a normal distribution out of BNFs. Previous works have utilized self-supervised pre-trained representations  \cite{10095191}  or ASR features (bottleneck or logits)\cite{10094737} as content encoder outputs. These features contain information related to content, speaker, and accent. To filter out unwanted information, the \(BNFs\) are processed by the bottleneck extractor. The TTS model's prior encoder consists of a text encoder \(\theta_{text}\) , which takes the transcript as input. Monotonic Alignment Search (MAS) aligns the text encoder output with the normalizing flow output, to match the differing lengths for text representations and audio representations.

The training loss is split into  Variational Auto-encoder (VAE) loss and Generative Adversarial Network (GAN) loss. GAN loss includes adversarial loss \(L_{adv}(D)\) and \(L_{adv}(G)\) for the discriminator and generator, along with feature matching loss \(L_{fm}(D)\) for the generator \(G\) . VAE loss consists of reconstruction loss \(L_{rec}\) (L1 distance between target and predicted mel-spectrogram) and \(L_{kl}\) (KL divergence between the prior and the posterior). To address training-inference mismatch of VITS, we integrate the full end-to-end inference during training to improve voice quality. Specifically, L1 distance between ground-truth  \(x_{mel}\) and inference output \({x}^{e2e}_{mel}\) generated by \(p_\psi(x| f^{-1}(z^{'}), g, F0)\), where the latent \(z^{'}\) is sampled from \(p_{\theta_{audio}}(z^{'}|c_{\text{audio}})\) or \(p_{\theta_{text}}(z^{'}|c_{\text{text}})\), and both \(g\) and \(F0\) are extracted from speaker encoder and F0 encoder respectively,  is added to reconstruction loss.  KL loss \(L_{kl}\), which sums the KL divergence between the prior distributions \(p_{\theta_{text}}(z|c_{\text{text}})\) and \(p_{\theta_{audio}}(z|c_{\text{audio}})\), and the posterior distribution \(q_\phi(z|x)\). The overall training loss for pre-training phase can be expressed:
\vspace{-0.1cm}
\begin{equation}
L_{kl} = {KL}(q_\phi(z|x)||p_{\theta_{text}}(z|c_{text})) + {KL}(q_\phi(z|x)||p_{\theta_{audio}}(z|c_{audio}))
\label{eq2}
\end{equation}
\begin{equation}
L_{recon} = || x_{mel} - {\hat{x}}_{mel} || + || x_{mel}- {x}^{e2e}_{mel}|| 
\label{eq3}
\end{equation}
\begin{equation}
L(G) = L_{rec} + L_{kl} + L_{adv}(G) + L_{fm}(G)
\end{equation}
\begin{equation}
L(D) = L_{adv}(D)
\end{equation}

\subsection{Generating ideal ground-truth}\label{generate}
The native TTS model trained earlier is used to generate ground-truth audio for each non-native input. We start by sampling the latent variable \(z\) from the posterior distribution of the non-native audio \(x\) . We use MAS to find the alignment \(A\) to maximize the log-likelihood of \(f(z)\) on the distribution \(p_{\theta_{text}}(z^{'}|c_{text},A)\) , which is used to upsample the text input.
\begin{equation}
z \sim {q_\phi(z|x_{non-native})}=N(z;\mu_\phi(x),\sigma_\phi(x))
\label{eq5}
\end{equation}
\begin{equation}
A = arg max  \log {N}({f(z)}; \mu_{\theta_{text}}(c_{text}, A), \sigma_{\theta_{text}}(c_{text}, A))
\label{eq6}
\end{equation}
\begin{equation}
c_{upsample} = upsampling(c_{text},A)
\label{eq7}
\end{equation}

Next, similar to the inference process of VITS, involves obtaining the latent variable \(z^{'}\) by sampling from  \(p_{\theta_{text}}(z^{'}|c_{upsample})\), which is then refined through an inverted normalizing flow \(f^{-1}\). Finally, the ground-truth audio \(\hat{y}_{\text{ground-truth}}\) is generated from \(p_\psi(x| f^{-1}(z^{'}), g, F0)\) using the HiFi-GAN decoder, where speaker embedding \(g\) and \(F0\) are extracted from non-native audio. In summary, the synthetic ground truth \({y}_{ground-truth}\) 
is generated from the transcripts with perfect native pronunciation, while utilizing the MAS alignment, \(F0\) and \(g\) from the non-native audio to retain the original duration, prosody and speaker identity.
\begin{equation}
z^{'} \sim p_{\theta_{text}}(z^{'}|c_{upsample})
\label{eq8}
\end{equation}
\begin{equation}
y_{ground-truth} = {Hifigan( f^{-1}(z^{'} ), g, F0)}
\label{eq9}
\end{equation}
\subsection{Finetuning AC model with ideal ground-truth and knowledge distillation from native TTS}

In the second stage, we continue fine-tuning the AC model using synthetic ground-truth data. The primary objective is to enable the model to generate ideal native output from non-native input. Specifically, the content encoder processes the non-native input, while other components take the synthetic ground-truth, as illustrated in Figure 1b. During this process, we freeze the parameters of all components except the bottleneck extractor and HiFi-GAN decoder. The bottleneck extractor is fine-tuned to capture accent-independent content representations from non-native inputs. To facilitate this, we introduce a distillation loss into the training, which is the KL divergence between the two prior distributions \(p_{\theta_{text}}(z|c_{\text{text}})\) and \(p_{\theta_{audio}}(z|c_{\text{audio}})\). The training loss during the fine-tuning stage is defined as:
\begin{equation}
L_{distill} = {KL}(p_{\theta_{audio}}(z|c_{audio})||p_{\theta_{text}}(z|c_{text}))
\end{equation}
\begin{equation}
L(G) = L_{rec} + L_{kl} + L_{distill} + L_{adv}(G) + L_{fm}(G) 
\end{equation}
\subsection{Overall model architecture detail}
Our posterior encoder, HiFi-GAN decoder, normalizing flow, and text prior encoder follow the original VITS framework \cite{kim2021conditional}. For the content encoder, we fine-tuned a pretrained Wav2vec 2.0 model on the Librispeech and L2Arctic dataset using CMU-dict phoneme labels, utilizing the bottleneck of the final layer as the content representation. Afterward, the content encoder's parameters are frozen. To get better text-audio alignment, we input phoneme labels instead of text into the text encoder. Both bottleneck extractor and text encoder are the same as the VITS Transformer-based text encoder. We retain Wav2vec 2.0's downsampling rate of 320 for calculating mel-spectrograms, F0, and linear spectrograms, while the HiFi-GAN decoder uses a corresponding upsample rate of 320. F0 sequences are derived using the YAAPT algorithm with the same downsampling rate, while the F0 encoder is based on \cite{wang21n_interspeech}. We employ the pre-trained speaker encoder from \cite{10.1109/TASLP.2021.3076867}.

\subsection{Inference}
\label{sec:format}
Our framework supports inference with or without a transcript. Without a transcript, the model extracts content information from non-native audio using content encoder and the bottleneck extractor in the prior encoder, then generates a waveform from the latent variable sampled from the prior distribution, conditioned on the speaker and prosody embedding. When a transcript is provided during inference, the system can utilize the prior text encoder instead of the audio encoder, as described in Section \ref{generate}, to genarate output with native pronunciation. This dual inference approach enhances inference quality, particularly when a transcript is available.

\vspace{-0.2cm}
\section{Experiment and Result}
\label{sec:pagestyle}
\subsection{Data}
We use the LJspeech dataset \cite{ljspeech17}, which features consistent pronunciation from a single native speaker.  Using FreeVC \cite{10095191}, a voice conversion model, we generate multispeaker native-accented utterances from the original voices. This augmented multi-speaker dataset is utilized, enabling the system to be trained in a multi-speaker setting during the first stage. In the second stage, we fine-tune the AC model on the L2ARCTIC dataset \cite{zhao18b_interspeech}, featuring 24 accented speakers across 6 accents. For each accent, we select 3 speakers for training and the remaining speakers for testing. Each speaker's utterances are split into a training set of 1,032 non overlap utterances, a validation set of 50, and a test set of 50. The test set, chosen using our competive ASR model \cite{pham20_interspeech}, focuses on utterances with the high average WER (larger than 10) across all speakers, with assumption that the higher WER rates mean stronger accents. 
\vspace{-2mm}
\subsection{Evaluation metrics}

\subsubsection{Subjective tests}

\textbf{Nativeness and Speaker Similarity Tests:} Ten participants conducted two evaluations using a 5-point scale (1-bad, 2-poor, 3-fair, 4-good, 5-excellent). In the Nativeness test, they rated how closely the converted audios resembled native speech. In the speaker similarity test, participants assessed the similarity between the voice identity of the original input and the converted audio, providing a score out of 5 for each.

\subsubsection{Objective tests}

\textbf{Word error rates (WER), Accent classifier accuracy (ACC) and Speaker Embedding Cosine Similarity (SECS) :} To evaluate the improvement in pronunciation, we use the Word Error Rate (WER) from our competitive seq2seq Transformer ASR model \cite{pham20_interspeech}. A lower WER indicates better intelligibility in the conversion process. SECS is employed to measure speaker similarity by computing the cosine similarity between speaker embeddings of the original speech and the converted speech. We utilize the state-of-the-art speaker verification model from \cite{9814838}  to extract speaker embeddings, which suggests that a cosine similarity greater than 0.85 indicates that both audios likely originate from the same speaker.

Additionally, we train an accent classifier to determine whether an input audio is native or non-native. The classifier architecture and training setting is similar to \cite{AC_NamNguyen}. We compute two accuracy scores, one for original non-native audio set and one for converted native audio set. A better AC is expected to show a larger gap between these two ACCs. 
\vspace{-0.2cm}

\subsection{Experimental setup}
Trained on both native and non-native audio with ASR loss, the content encoder's final layer yields representations closely aligned with linguistic content. This makes it more accent- and speaker-independent, facilitating accurate mapping from non-native pronunciation to the correct linguistic information. Therefore, we use the model which is pre-trained in the first stage as the baseline. The proposed model is the model fine-tuned in the second stage. To assess the effectiveness of fine-tuning on the synthetic ground-truth data, we train a variant that only uses knowledge distillation loss \(L_{distill}\) between the text and audio encoders in the second stage. Additionally, the audio quality of the Synthetic ground-truth is also evaluated, allowing us to assess the system's performance when the correct transcript is provided. The training hyperparameters are set similarly to those in the original VITS, with the system trained for 600,000 steps in the first stage and 200,000 steps in the second stage. Sample evaluation audios are available at a github repository \footnote{\url{https://accentconversion.github.io/demo}}

\subsection{Result}
The objective metric evaluation, shown in Table \ref{tab:result2}, indicates that the proposed method outperforms others in terms of WER. This demonstrates that the method successfully improves pronunciation, making it closer to native speech.  Even without synthetic ground-truth, the model still benefits from knowledge distillation from the native TTS, although the pronunciation improvement is smaller. The synthetic ground -truth, generated from transcripts, has a WER of 5.1, confirming the high quality of these ground-truth audios. Additionally, SECS remains stable between 0.82 and 0.84 across all settings, verifying that the speaker identity is well-preserved in all methods.  In terms of ACC, the results are not as strong as other metrics. Since we aim to preserve the prosody of the original audio, which can sometimes indicate the accent of the speaker, the model occasionally identifies the converted audio as non-native. This prosody retention may lead to less effective accent conversion in some cases. The subjective metric evaluation, also shown in Table \ref{tab:result1}
, indicates that our proposed method performs best in the nativeness test, while the Sim-MOS results, like in the objective evaluation, remain similar across all setups.
\vspace{-0.3cm}
\begin{table}[ht]

	\setlength{\tabcolsep}{4pt}

	\centering
 \caption{Subjective metrics}
	\begin{tabular}{lccc}
	 \cline{1-3}
       \textbf{Models} &  {Nativeness} & {Sim-MOS} \\ 
       
        \cline{1-3}
        Original  & \(1.67\pm0.05\) & - \\      
        Synthetic ground-truth & \(3.93\pm0.05\) & \(3.84 \pm0.15\)  \\
         \cline{1-3}         
         Baseline & \(3.32\pm0.08\) & \(3.94\pm0.18\)\\
         w/o Synthetic ground-truth & \(3.71\pm0.08\) & \(3.9\pm0.19\) \\
         Proposed & \(\textbf{3.87}\pm0.08\)  & \(3.85\pm0.19\)\\
         \cline{1-3}
	\end{tabular}

\label{tab:result1}
\vspace{-3mm}
\end{table}

\vspace{-0.3cm}
\begin{table}[ht]
        \centering
	\setlength{\tabcolsep}{4pt}
	
\caption{Objective metrics}

	\begin{tabular}{lcccc}
	 \cline{1-4}
       \textbf{Models} &  {WER} & {ACC} & {SECS}\\ 
       
        \cline{1-4}
        Original  & 18.3 & 98.3 & 1.0 &\\
      
        Synthetic ground-truth & 5.1 & 13.9 & 0.83  \\
         \cline{1-4}
         
         Baseline & 17.1 & 18.9 & 0.82 \\
         w/o Synthetic ground-truth & 14.3  & 17.0 &  0.84\\
         Proposed & \textbf{12.4} &  17.2 & 0.83 \\
         \cline{1-4}
	\end{tabular}

\label{tab:result2}
\vspace{-3mm}
\end{table}

\section{Conclusion}
Our work presents a significant advancement in accent conversion and native-like pronunciation correction. Through the introduction of a two-stage training process that leverages VITS framework,  we enable the AC model to better capture the accent-independent feature of speech while maintaining the speaker’s identity. The use of generated ground-truth data, alongside knowledge distillation from native TTS, further enhances the system’s ability to correct non-native pronunciations effectively. Objective and subjective evaluations demonstrate that our approach not only improves accent conversion but also addresses pronunciation issues, providing a more natural and comprehensible output. 
\label{sec:typestyle}

\section{Acknowledgment}
This research was supported by a grant from Zoom Video Communications , Inc. T;  European Commission Project Meetween (101135798), the Federal Ministry of Education and Research (BMBF)
 of Germany under the number 01EF1803B (RELATER), and the pilot program Core-Informatics of the Helmholtz Association (HGF). The work is enabled by the Horeka supercomputer (BMBF) and the DeltaGPU at NCSA/ACCESS grant\cite{boerner2023access}.

\bibliographystyle{IEEEtran}
\bibliography{ref}

\end{document}